\documentclass[a4paper,11pt]{article}
\usepackage{pos}
\usepackage{caption}
\usepackage{subfigure}
\usepackage{xspace}
\usepackage{lineno}

\newcommand{ \pt}{\ensuremath{p_\mathrm{T}}\xspace}
\newcommand{ \pttrig}{\ensuremath{p_\mathrm{T}^\mathrm{trigg}}\xspace}
\newcommand{ \ptassoc}{\ensuremath{p_\mathrm{T}^\mathrm{assoc}}\xspace}
\newcommand{ \dphi}{\ensuremath{\Delta\varphi}\xspace}
\newcommand{ \deta}{\ensuremath{\Delta\eta}\xspace}
\newcommand{ \K}{\ensuremath{\mathrm{K_S^0}}\xspace}
\newcommand{ \lamAlam}{\ensuremath{(\Lambda+\overline{\Lambda})}\xspace}
\newcommand{ \lam}{\ensuremath{\Lambda}\xspace}
\newcommand{ \Alam}{\ensuremath{\overline{\Lambda}}\xspace}
\newcommand{ \vzero}{\ensuremath{\mathrm{V}^0}\xspace}
\newcommand{\gevc}{\ensuremath{\mathrm{GeV/}c}\xspace}

\title{Exploring jet fragmentation using two-particle correlations with $\Lambda$ and K$^0_{\rm S}$ }

\author*[a]{Lucia Anna Tarasovi\v cov\'a}

\affiliation[a]{Westfälische Wilhelms-Universität,\\
  Wilhelm-Klemm-Straße 9, Münster, Germany}

\emailAdd{lucia.anna.husova@cern.ch}

\abstract{
Complementary to jet reconstruction, two-particle correlations in $\Delta\eta$ and $\Delta\varphi$ are used to study jets, in particular their particle composition. While in Pb--Pb collisions this is done to characterize the quark--gluon plasma, pp and p--Pb collisions serve as a reference and are of interest on their own for their input into the understanding of particle production mechanisms. Recent ALICE results on the production of strange particles in small systems (pp and p--Pb collisions) reveal the possibility of having similar strange hadron production mechanisms in all collision systems. We present here a study of two-particle correlations triggered with strange hadrons ($\mathrm{K_S^0}$, $\Lambda$, $\overline{\Lambda}$) in pp collisions at 13 TeV and 5.02 TeV and in the most central Pb--Pb collisions at $\sqrt{s_{\mathrm{NN}}}$= 5.02 TeV. The dependence of the per-trigger yields of primary charged hadrons on the transverse momenta of the trigger and associated particles, as well as on the event multiplicity, will be presented for both the near-side and away-side regions. Moreover, the ratios of these yields to the yields extracted from inclusive hadron-hadron correlations and the nuclear modification factor $I_{AA}$ will be discussed. The results are compared among the three hadron species. In addition, a comparison to different Monte Carlo generators is presented, which allows us to better understand strangeness production in jets.
}

\FullConference{%
  *** The European Physical Society Conference on High Energy Physics (EPS-HEP2021), ***\\
  *** 26-30 July 2021 ***\\
  *** Online conference, jointly organized by Universität Hamburg and the research center DESY ***
}

\begin{document}
\maketitle

\section{Introduction}

Jet properties depend strongly on the original parton. These differences between quark and gluon jets were observed in $e^++e^-$ annihilations at LEP by the OPAL and DELPHI collaborations~\cite{OPAL_qg}, \cite{DELPHI_qg_mult}, \cite{OPAL_qg_mult}.  A fragmenting gluon emits more soft gluons than a fragmenting quark which leads to more particles and a wider cone in the case of gluon jets.  Moreover, the OPAL experiment measured the relative production of \K mesons and \lam baryons to the total charged particle production in quark and gluon jets and observed an enhanced relative production of \lam in gluon-driven jets~\cite{OPAL_klam}. Nevertheless, the jet tagging based on the original parton (quark or gluon) is much more difficult in pp and in Pb--Pb collisions where the differentiation between them would help to study the interaction between quarks, gluons and QGP.

In order to study jets, the two-particle correlation method is used here in pp collisions at 13 TeV and pp and Pb--Pb collisions at $\sqrt{s_{\mathrm{NN}}}$ = 5.02 TeV.  The correlations are triggered with strange \vzero hadrons (\K, \lam, \Alam) and charged primary hadrons (h), dominated by pions. The selection of a (relatively) high transverse momentum provides a proxy for the jet axis. Thus, the trigger particles are selected with following requirements: 3 $<$ \pttrig $<$ 20 \gevc and 8 $<$ \pttrig $<$ 16 \gevc for the collision energy 13 TeV and 5.02 TeV, respectively. These are associated with other charged hadrons with lower transverse momentum (1 \gevc $<$ \ptassoc $<$ \pttrig). In order to build a  correlation function, angular differences between trigger and associated particles are calculated in the \dphi, \deta space.  Such two-dimensional correlation function is normalised per trigger particle., projected on the \dphi axis and the underlying event is subtracted.  The latter is assumed to be flat in pp collisions and modulated due to collective flow in Pb--Pb collisions.  Two peaks remain, the near- and away-side,  and by integrating them the per-trigger yields can be extracted. The effect of the QGP on these peaks can be studied by yield ratios in big and small collisions systems and the different jet fragmentation processes can be investigated by means of the ratios of yields triggered with different hadron species.

Because of the neutrality of the \vzero hadrons, they need to be reconstructed from their decays into charged particles: $\K \rightarrow \pi^+ + \pi^-$ (69\%), $\lam \rightarrow p+\pi^-$ (64\%) and $\Alam \rightarrow \overline{p}+\pi^+$ (64\%)~\cite{pdg}. The combinatorial background is suppressed by applying selection criteria based on the decay topology and the daughter tracks are identified through their specific energy loss in the TPC. The residual background is subtracted with the help of side-bands of the invariant mass spectrum.  

\section{Results}

\begin{figure}[hbtp!]
\centering
\includegraphics[width=15cm]{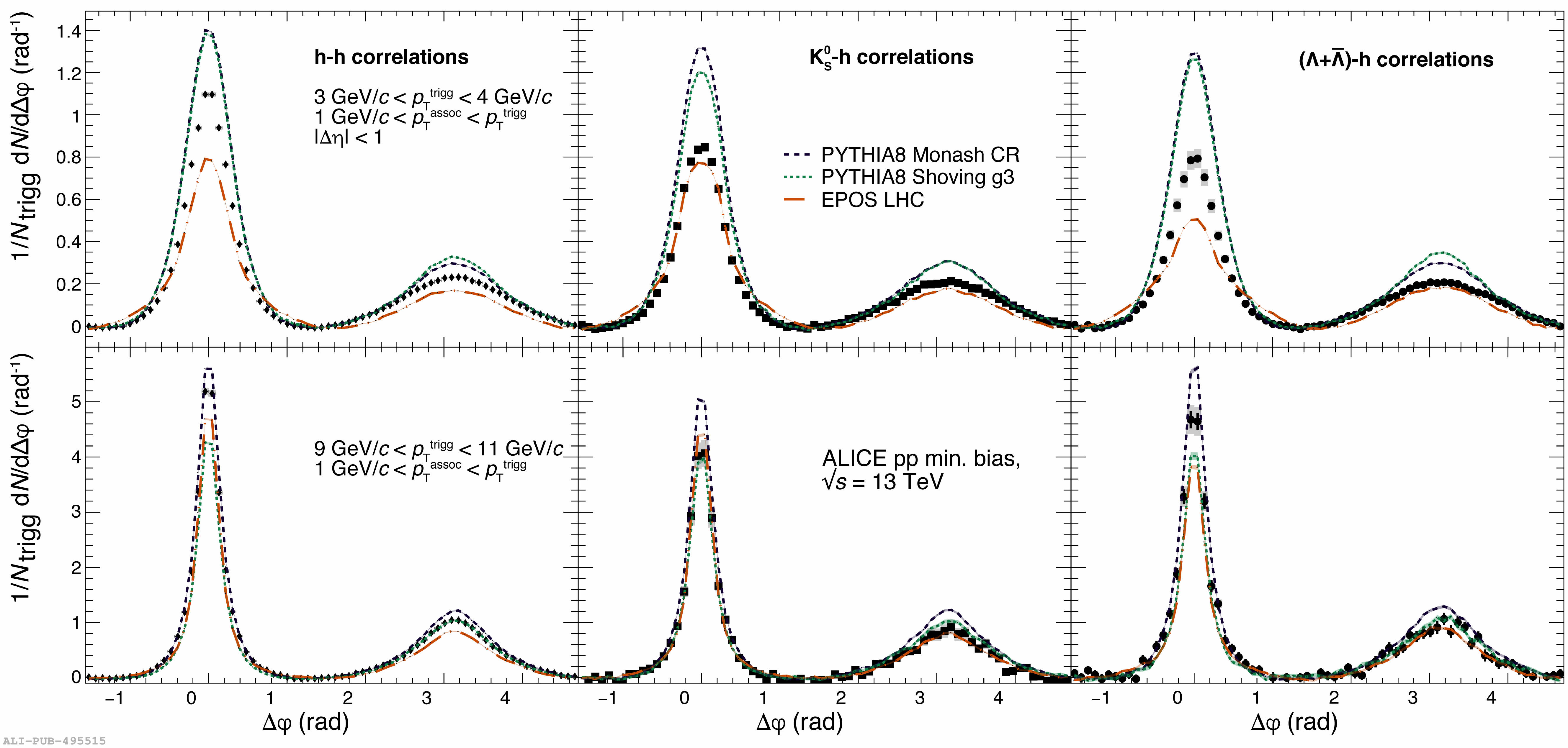}
\caption{\dphi projections of h-h (left), \K -h (middle) and \lamAlam -h (right) correlation functions for two \pttrig intervals, 3 $<$ \pttrig $<$ 4 \gevc (top) and 9 $<$ \pttrig $<$ 11 \gevc (bottom) compared with MC models. }
\label{DeltaPhi_pp}
\end{figure}

In Fig.~\ref{DeltaPhi_pp}, the \dphi projections of the three types of studied correlation functions are shown for the pp collisions at 13 TeV~\cite{our}.  Each of the projections is compared with three Monte Carlo (MC) models - PYTHIA8 Monash~\cite{pythia_mon}, PYTHIA8 with shoving~\cite{pythia_shov} and EPOS LHC~\cite{epos}.  None of the models is able to describe the magnitude of the correlation function for all three types of trigger particles in both shown \pttrig intervals.  EPOS  LHC fails in the description in all cases except for the \K -h correlation function while PYTHIA8 Monash is closer to data in the higher \pttrig interval.  However, there is almost no difference between the two considered PYTHIA8 models in low \pt, a difference can be observed in the higher \pttrig interval where the shoving model underestimates the data strongly.  

\begin{figure}[hbtp!]
\centering
\begin{subfigure}
		\centering
         \includegraphics[width=15cm]{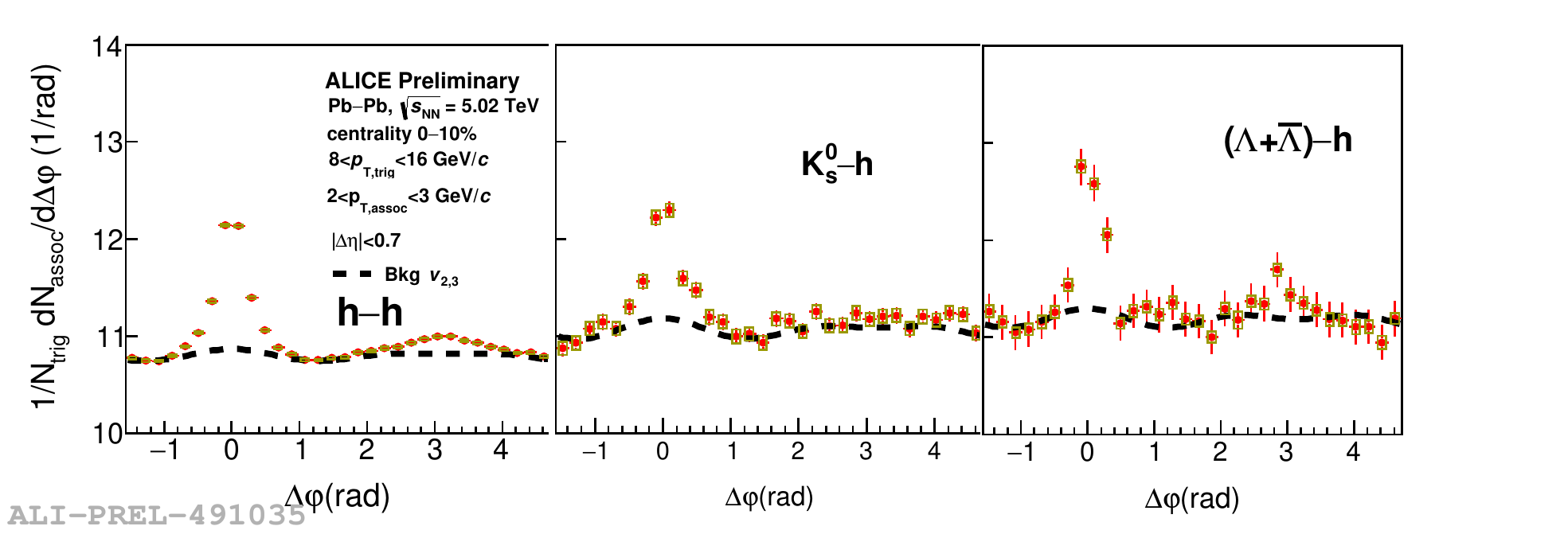}
\end{subfigure}         
\begin{subfigure}
		\centering
         \includegraphics[width=15cm]{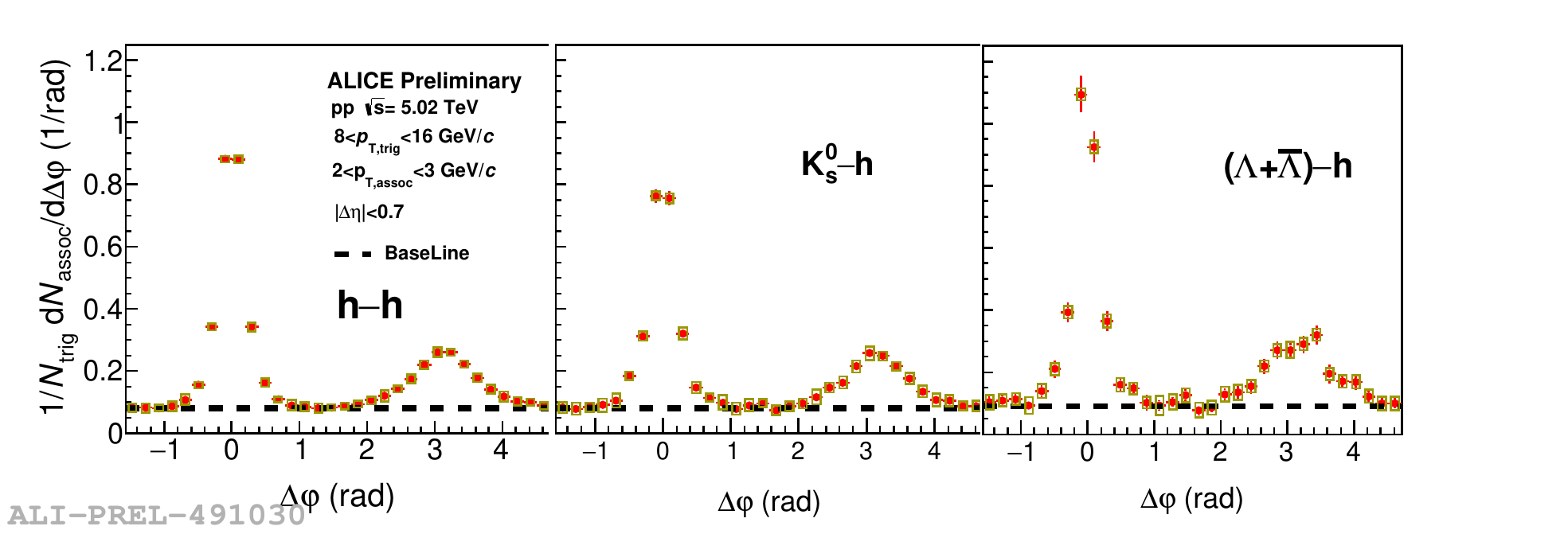}
\end{subfigure}    
\caption{\dphi projections of h-h (left),  \K -h (middle) and \lamAlam -h (right) in central (0-10\%) Pb--Pb (top) and pp (bottom) collisions for the centre-of-mass energy 5.02 TeV with marked underlying event contribution.}
\label{DeltaPhi_PbPb}
\end{figure}

The \dphi projections of the same types of correlation functions for collision energy 5.02 TeV are plotted in Fig.~\ref{DeltaPhi_PbPb} for both pp and central (0-10\%) Pb--Pb collisions. The modulation of the magnitudes of the peaks can be seen: the yield of the near-side peaks is enhanced for the correlation functions measured in Pb--Pb collisions while the yields of away-side peaks are suppressed due to parton energy loss in the QGP. 

\subsection{Per-trigger yields}

\begin{figure}[hbtp!]
\centering
\includegraphics[width=15cm]{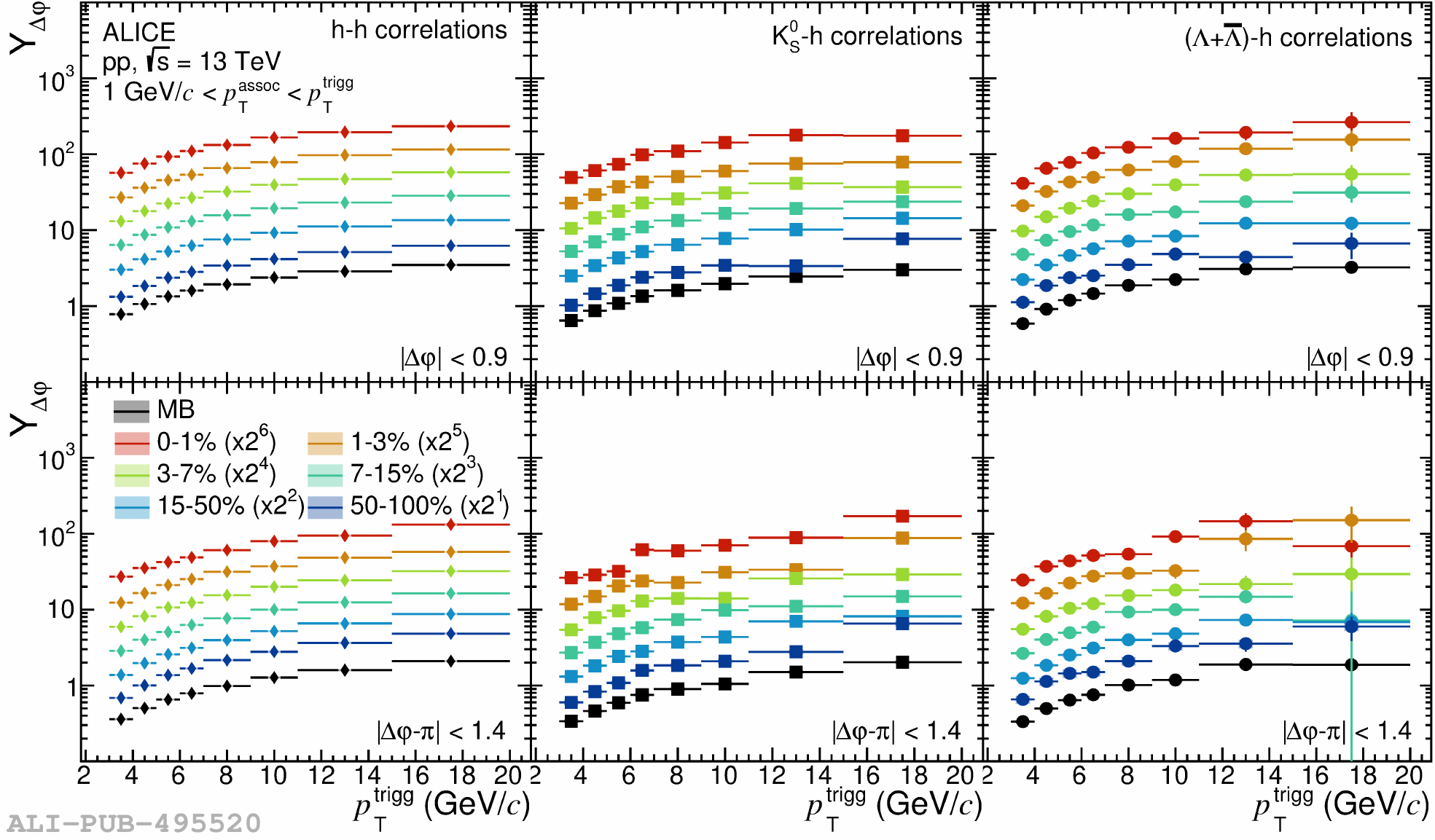}
\caption{Per-trigger yields from h-h (left), \K -h (middle) and \lamAlam -h (right) correlation functions on the near-side (top) and away-side (bottom) for different multiplicity classes in pp collisions at 13 TeV.}
\label{yields_pp}
\end{figure}

Fig.~\ref{yields_pp} is showing the per-trigger yields for different trigger particles as a function of \pttrig for different multiplicity classes. An increasing trend with \pt is observed. This is caused by the more available energy in a jet triggered with hadron with higher transverse momentum.  Quantitatively, the yields are similar for all trigger particles on both near- and away-side. 

\subsection{$I_{AA}$}

\begin{figure}[hbtp!]
\centering
	\begin{subfigure}
		\centering
		\includegraphics[width=6.7cm]{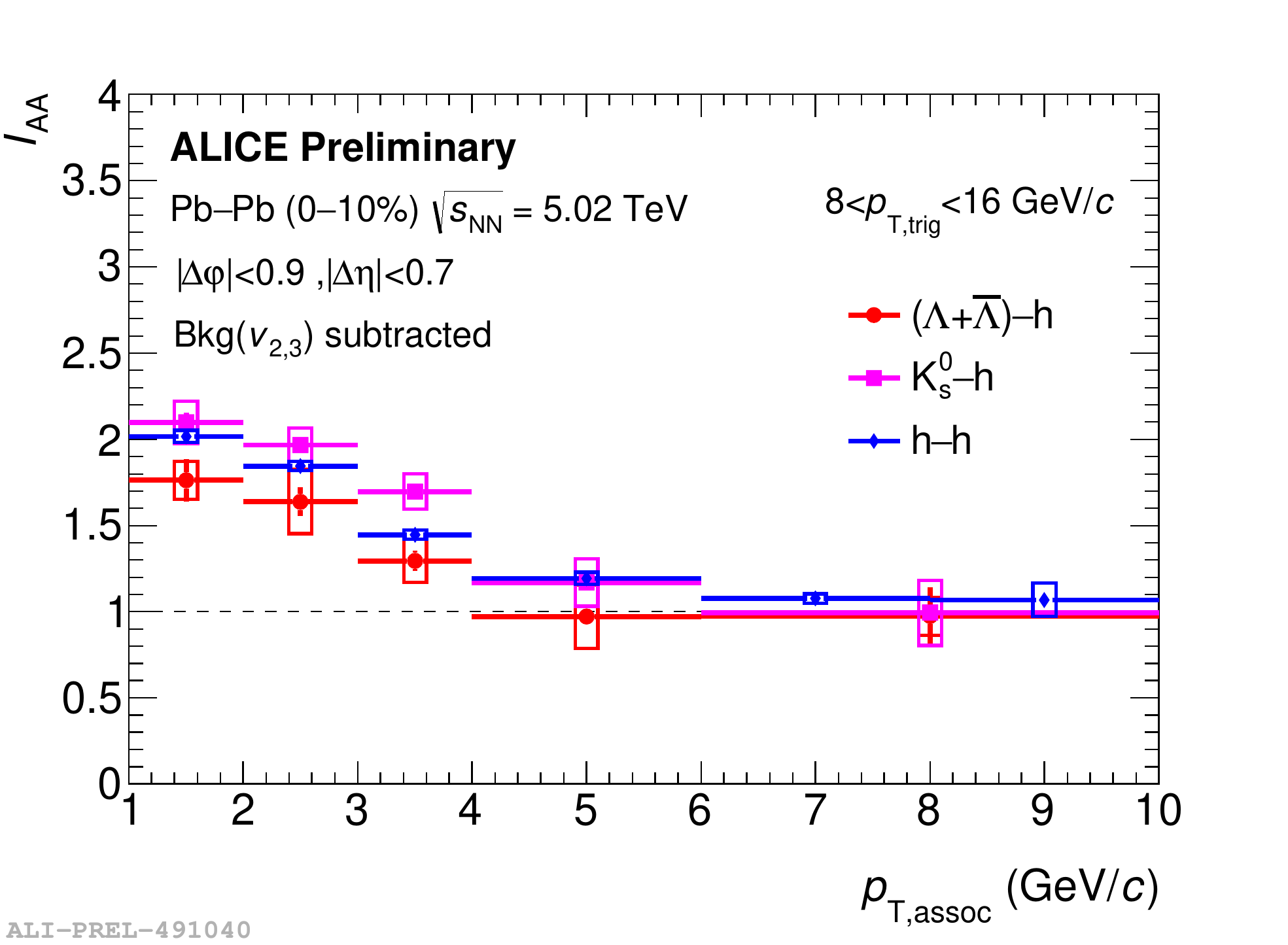}
	\end{subfigure}
	\begin{subfigure}
		\centering
		\includegraphics[width=6.7cm]{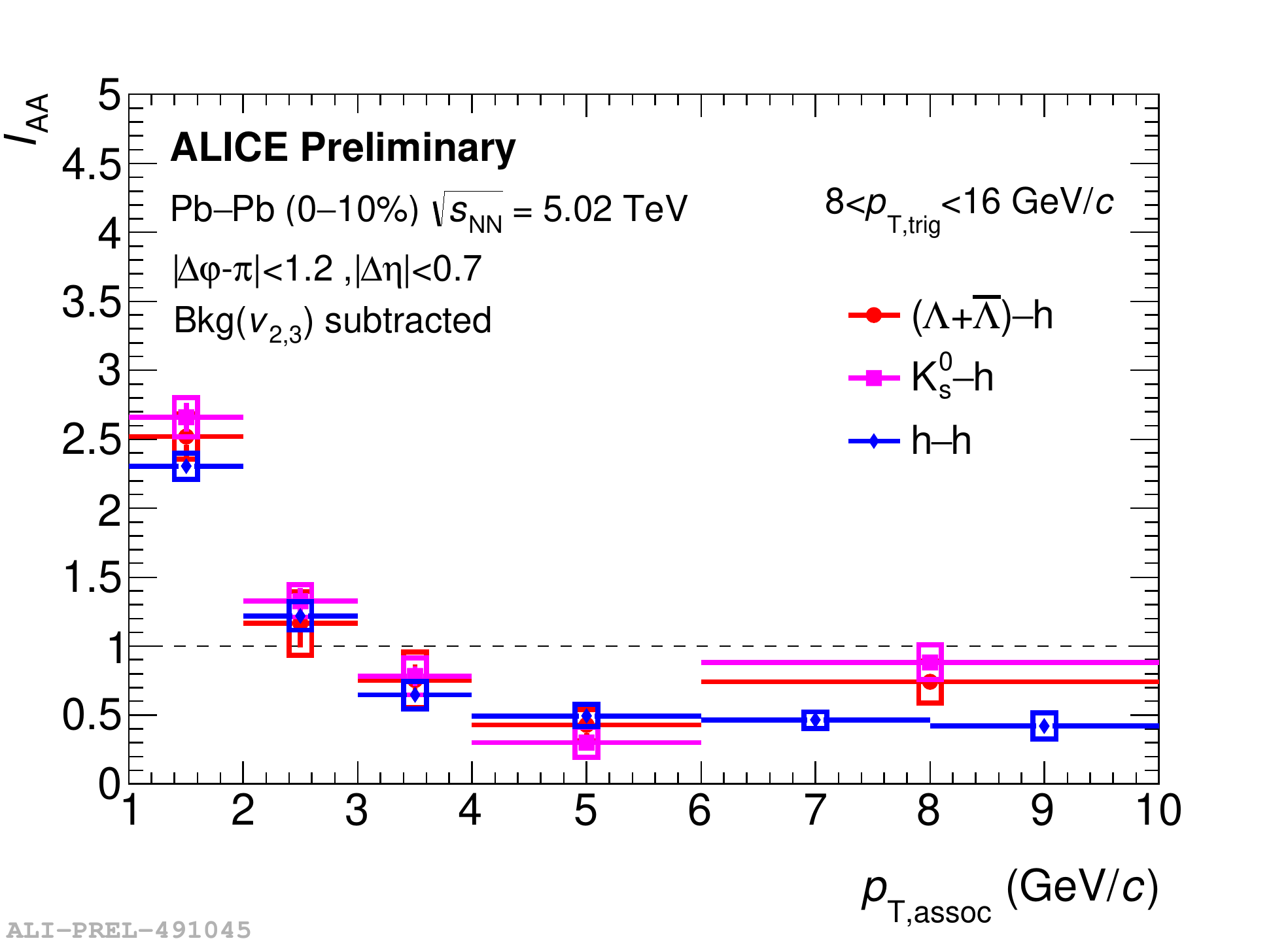}
	\end{subfigure}
\caption{Ratio of the per-trigger near-side (left) and away-side (right) yields in Pb--Pb collisions to the ones in pp collisions for different trigger particles.}
\label{Iaa}
\end{figure}

One of possible ways to study the impact of the QGP to the jet fragmentation is the calculation of the $I_{AA}$ ratio which is defined as $\mathrm{Y_{\dphi}^{Pb-Pb}}/\mathrm{Y_{\dphi}^{pp}}$.  This ratio is shown in Fig.~\ref{Iaa} where a strong enhancement is observed for low \ptassoc, both on the near- and away-side.  For high \ptassoc , no modification is observed on the near-side while there is a strong suppression on the away-side. This suppression is caused by the energy loss in the QGP.~\cite{yield_ALICE_PbPb}.  
The left panel of Fig.~\ref{Iaa} shows that the near side $I_{AA}$ is slightly depleted for the \lamAlam triggers with respect to the other trigger particles, but this effect is not significant.

\subsection{Comparison of hadron- and strangeness-triggered yields}

\begin{figure}[hbtp!]
\centering
	\begin{subfigure}
		\centering
		\includegraphics[width=6.7cm]{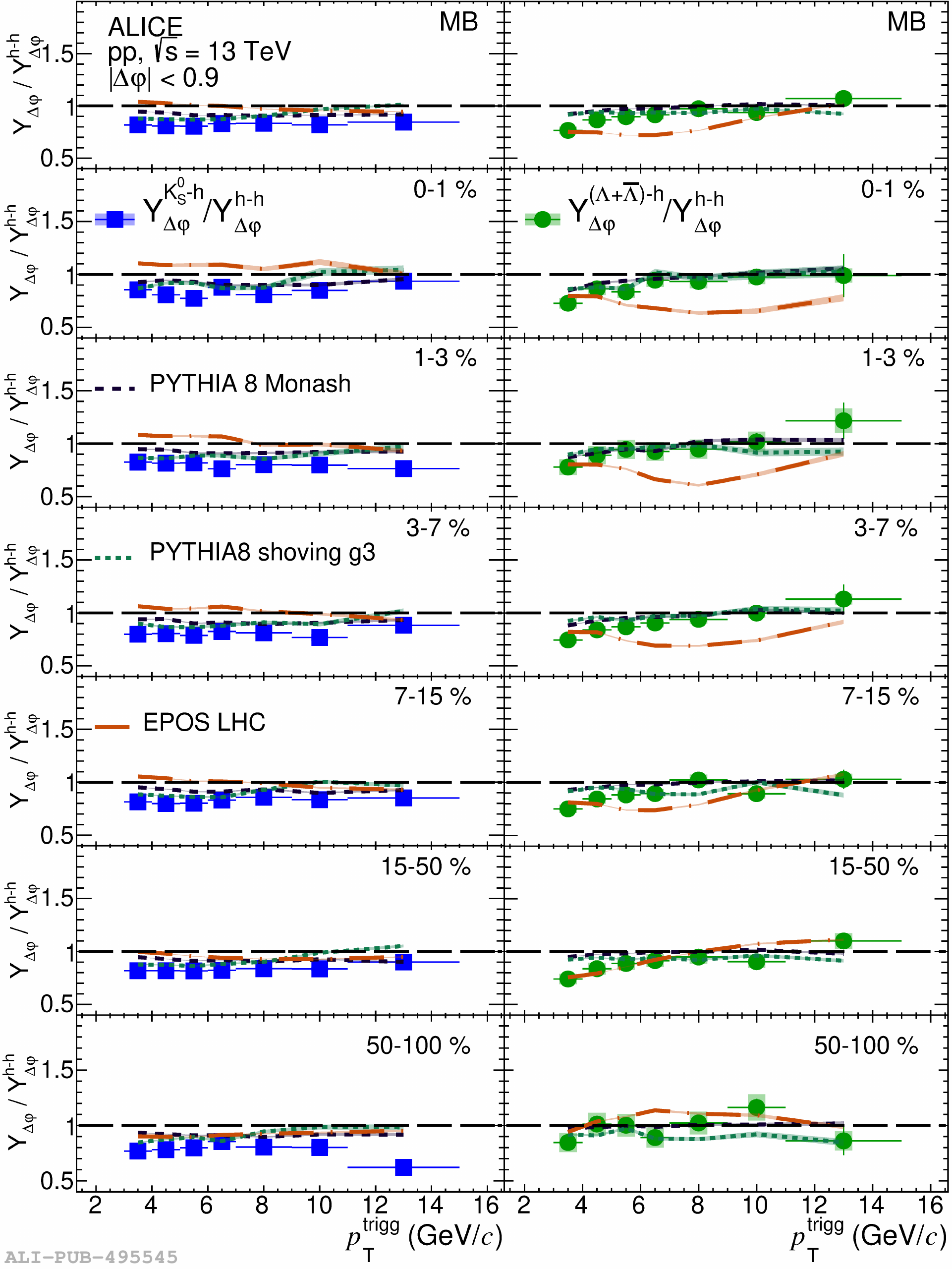}
	\end{subfigure}
	\begin{subfigure}
		\centering
		\includegraphics[width=6.7cm]{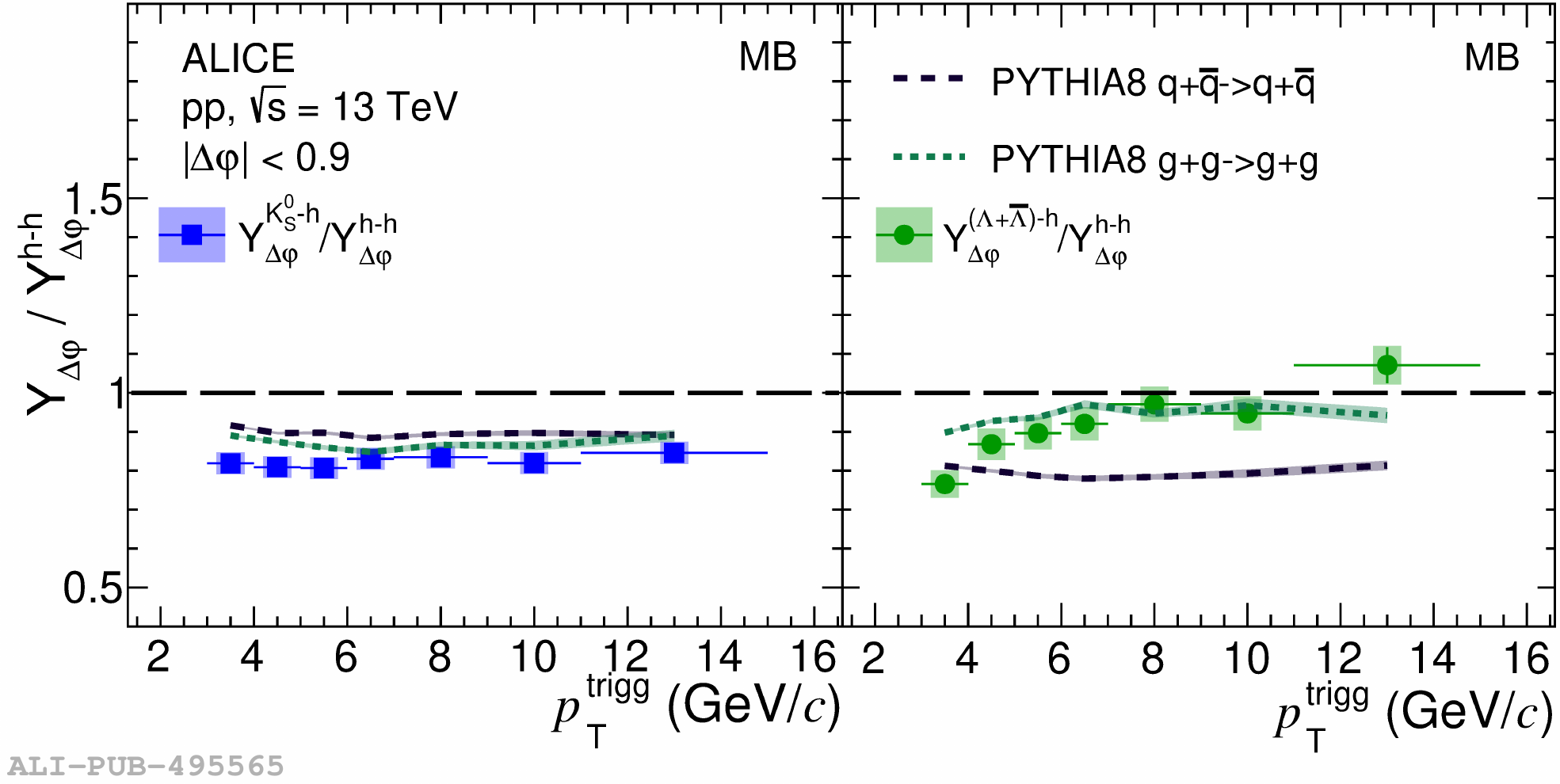}
	\end{subfigure}
\caption{Ratios of integrated per-trigger yield of \K -h (left column) or \lamAlam -h (right column) to h-h as a function 
of \pttrig for the near-side for different event multiplicity classes in the left plot and in the right plot MB results compared  with the PYTHIA8 calculation of the quark and gluon jets.}
\label{ratio_tohh}
\end{figure}

In order to be able to study the differences between different trigger particles in more details, the ratios of yields triggered with \K or \lamAlam over the ones extracted from h-h correlation functions are calculated. The results are shown in the left part part of  Fig.~\ref{ratio_tohh}.  Here, the ratios are compared with models which show better agreement with the data than it was for the correlation functions (Fig.~\ref{DeltaPhi_pp}). This is caused by similar mismatch between data and models for all trigger particles. In a detailed observation, one can see a difference between the ratios triggered with \K and \lamAlam. The $Y_{\dphi}^{\K-h}/Y_{\dphi}^{h-h}$ is rather flat with \pttrig and smaller than unity. This is probably due to the fact that jets triggered with a \K meson fragment in a smaller amount of charged particles with respect to the jets triggered with a pion and this effect does not depend on the initial energy of the trigger particle.  The distribution does not show no dependence on the event multiplicity. However, the ratio $Y_{\dphi}^{\lamAlam-h}/Y_{\dphi}^{h-h}$ increases with \pttrig which means that the jets triggered with high-\pt  \lam or \Alam have more associated particles than jet triggered with a pion with the same \pt. This could point out that triggering with high \pt \lam or \Alam hyperon biases the jet selection towards gluon jets which have higher particle multiplicity~\cite{OPAL_qg_mult},\cite{DELPHI_qg_mult}. To prove this assumption, a separate PYTHIA8 study was performed where pure gluon ($g+g\rightarrow g+g$) or quark ($q+\overline{q}\rightarrow q+\overline{q}$) hard processes were simulated. The comparison of results from this study with data in MB is shown in the right part of Fig.~\ref{ratio_tohh}.  There, it is visible that the ratios from both quark and gluon samples are almost the same in case of \K , but the ratio from gluon sample is shifted towards higher values and increases slightly with \pttrig in the case of \lamAlam. 

\section{Summary and conclusions}

The correlation functions triggered with charged hadrons, \K mesons and \lam (\Alam) hyperons are measured in pp and Pb--Pb collisions at 13 TeV and 5.02 TeV, with the ALICE experiment. The per-trigger yields are extracted and compared among collision systems and different trigger particles. The $I_{AA}$ ratio shows an enhancement for low \ptassoc on both near- and away side while a suppression on the away-side is visible for high \ptassoc . Both can be explained by the interaction of the initial jet parton with the QGP~\cite{yield_ALICE_PbPb}. The ratios between \vzero triggered and charged hadron triggered yields show a difference between \K and \lam(\Alam) being the trigger particle. This can be understood through a PYTHIA8 study as a bias towards gluon jets caused by triggering with \lam(\Alam).

\end{document}